\input{aipcheck}

\documentclass[
    ,final            % use final for the camera ready runs
  ]
  {aipproc}

\layoutstyle{8x11single}

\begin{document}

\title{$Q^2$-evolution of parton densities at small $x$
%low Q2
values\\
%Combined 
and H1 and
%$\&$
ZEUS experimental
%F2 
data.}


\author{A.V. Kotikov}{
%Laboratory of Theoretical Physics,
%Joint Institute for Nuclear Research, 141980 Dubna, Russia
}

\author{B.G. Shaikhatdenov}{
address={Joint Institute for Nuclear Research, 141980 Dubna, Russia}
%Laboratory of High Energy Physics,
%Joint Institute for Nuclear Research, 141980 Dubna, Russia
}

\begin{abstract}
It is shown that in 
the
leading twist approximation of the Wilson operator product expansion
with ``frozen'' and analytic 
strong coupling constants, considering the
%we show that
Bessel-inspired
behavior of 
the structure functions $F_2$ and 
the  derivative $\partial \ln F_2/\partial \ln(1/x)$
at small $x$ values,
obtained for
a flat initial condition in the DGLAP evolution equations,
leads to a
good agreement with the 
deep inelastic 
scattering  
H1 and ZEUS
experimental data from HERA.

\end{abstract}

\maketitle

%%%%%%%%%%%%%%%%%%%%%%%%%%%%%%%%%%%%%%%%%%%%
%% MAINMATTER
%%%%%%%%%%%%%%%%%%%%%%%%%%%%%%%%%%%%%%%%%%%%

\section{Introduction}

A reasonable agreement between HERA data  \cite{H197}-\cite{DIS02}
%\cite{H1ZEUS,H1slo,DIS02}
and the next-to-leading-order (NLO) approximation of
perturbative Quantum Chromodynamics (QCD)
has been observed for $Q^2 \geq 2$ GeV$^2$ (see reviews in \cite{CoDeRo}
and references therein), which gives us a reason to believe that
perturbative QCD is capable of describing the
evolution of the structure function (SF) $F_2$ and its derivatives
%structure functions
down to very low $Q^2$ values, where all the strong interactions
are conventionally considered to be soft processes.

A standard way to study the $x$ behavior of
quarks and gluons is to compare the data
with the numerical solution to the
%DGLAP
Dokshitzer-Gribov-Lipatov-Altarelli-Parisi (DGLAP)
equations
\cite{DGLAP}
by fitting the parameters of
$x$-profile of partons at some initial $Q_0^2$ and
the QCD energy scale $\Lambda$ \cite{Martin:2009bu,Ourfits}.
However, for the purpose of analyzing exclusively the
small-$x$ region, there is an alternative to carry out
%of doing 
a simpler analysis
by using some of the existing analytical solutions to DGLAP equations
in the small-$x$ limit \cite{BF1}--\cite{HT}.

To improve the analysis at low $Q^2$ values, it is important to consider
%It is important to extend the analysis of \cite{KoPa02} to low $Q^2$ range
%with a help of
the well-known infrared modifications of the strong coupling
constant. We will use its
%the 
``frozen'' and analytic versions (see,
\cite{Kotikov:2010bm,Cvetic:2009kw} and references therein).
%and \cite{ShiSo}, respectively).

\section{
Generalized doubled asymptotic scaling
%DAS
approach} \indent

At low-$x$ values
there is 
%the alternative of doing a simpler 
the simple 
%analysis by using some of the existing 
analytical solution of DGLAP evolution
%in the %small low-$x$ limit 
\cite{BF1}:
%--\cite{HT}:
%This was done so in \cite{BF1} where it was pointed out that 
the HERA small-$x$ data can be
interpreted in 
terms of the so-called doubled asymptotic scaling (DAS) phenomenon
related to the asymptotic 
behavior of the DGLAP evolution 
discovered many years ago \cite{Rujula}.

The original study of \cite{BF1} was extended in \cite{Munich,Q2evo,HT}
to include the finite parts of anomalous dimensions
of Wilson operators
% and Wilson coefficients
\footnote{ 
In the standard DAS approximation \cite{Rujula} only the singular
parts of the anomalous dimensions were used.}.
This has led to predictions \cite{Q2evo,HT} of the small-$x$ asymptotic 
%PDF
form 
of parton distribution functions (PDFs)
in the framework of the DGLAP dynamics
%equation 
starting at some $Q^2_0$ with
the flat function
 \begin{eqnarray}
f_a (Q^2_0) ~=~
A_a ~~~~(\mbox{hereafter } a=q,g), \label{1}
 \end{eqnarray}
where $f_a$ are the parton distributions multiplied by $x$
and $A_a$ are unknown parameters to be determined from the data.

%From now on, we 
We refer to the approach of \cite{Munich,Q2evo,HT} as
{\it generalized} DAS approximation. In that approach
%generalized DAS 
the flat initial conditions in Eq. (\ref{1}) determine the
basic role of the singular parts of anomalous dimensions,
as in the standard DAS case, while
the contribution from finite parts 
of anomalous dimensions and from Wilson coefficients can be
considered as corrections which are, however, important for better 
agreement with experimental data.
In the present paper, similary to
\cite{BF1}--\cite{HT}, we neglect
the contribution from the non-singlet quark component.

The flat initial condition (\ref{1}) corresponds to the case when parton density
%distributions
tend  to some constant value at $x \to 0$ and at some initial value $Q^2_0$.
%(\ref{1}).
The main ingredients of the results \cite{Q2evo,HT}, are:
\begin{itemize}
\item
Both, the gluon and quark singlet densities are presented in terms of two
components ($"+"$ and $"-"$) which are obtained from the analytic 
$Q^2$-dependent expressions of the corresponding ($"+"$ and $"-"$) PDF
%parton distributions 
moments.
\footnote{Such an approach has been developed  \cite{Albino:2011si}
recently also for the fragmentation function, 
whose first moments (ie mean multiplicities of quarks and gluons) were analyzed
\cite{Bolzoni:2012ii}. The results are 
in good agreement with the experimental data.}
%(see contribution \cite{Bolzoni:2012cv} by Paolo Bolzoni to this Proceedings).}
\item
The twist-two part of the $"-"$ component is constant at small $x$ at any 
values of $Q^2$,
whereas the one of the $"+"$ component grows at $Q^2 \geq Q^2_0$ as
\begin{equation}
\sim e^{\sigma},~~~
%\exp{\sigma},~~~
\sigma = 2\sqrt{\left[ \left|\hat{d}_+\right| s
%\ln \left( \frac{a_s(Q^2_0)}{a_s(Q^2)} \right) 
- \left( \hat{d}_{++}
%\hat{D}_+ 
+  \left|\hat{d}_+\right|
%\hat{d}_+ 
\frac{\beta_1}{\beta_0} \right) p
% \Bigl( a_s(Q^2_0) - a_s(Q^2) \Bigr)
\right] \ln \left( \frac{1}{x} \right)}  \ ,~~~ \rho=\frac{\sigma}{2\ln(1/x)} \ ,
\label{intro:1}
\end{equation}
where $\sigma$ and $\rho$
%$=\sigma/(2\ln(1/x))$ 
are the generalized Ball--Forte
variables,
\begin{equation}
s=\ln \left( \frac{a_s(Q^2_0)}{a_s(Q^2)} \right),~~
p= a_s(Q^2_0) - a_s(Q^2),~~~
\hat{d}_+ = - \frac{12}{\beta_0},~~~
%\hat{D}_+ 
\hat{d}_{++} =  \frac{412}{27\beta_0}.
\label{intro:1a}
\end{equation}
\end{itemize}
Hereafter we use the notation
$a_s=\alpha_s/(4\pi)$.
The first two coefficients of the QCD $\beta$-function in the 
${\overline{\mbox{MS}}}$-scheme
are $\beta_0 = 11 -(2/3) f$
%$(11/3) C_A - (4/3) T_R f$ 
and $\beta_1 =  102 -(114/9) f$
%$(2/3)[17 C_A^2 - 10 C_A T_R f - 6 C_F T_R f]$ 
with $f$ is being the number of active quark flavors.

Note here that the perturbative coupling constant $a_s(Q^2)$ is different at
the leading-order (LO) and NLO approximations. Indeed, from the renormalization group equation
we can obtain the following equations for the coupling constant
%\begin{subequations}
%\label{as:LO&NLO}
\begin{eqnarray}
 \frac{1}{a_s^{\rm LO}(Q^2)} \, = \, \beta_0 
 \ln{\left(\frac{Q^2}{\Lambda^2_{\rm LO}}\right)},~~~~~
\frac{1}{a_s(Q^2)} \, + \,
 \frac{\beta_1}{\beta_0} \ln{\left[
 \frac{\beta_0^2 a_s(Q^2)}{\beta_0+ \beta_1 a_s(Q^2)}\right]} \, = \, 
 \beta_0 \ln{\left(\frac{Q^2}{\Lambda^2}\right)}
\label{as:LO} 
\end{eqnarray}
at the LO and NLO approximations, respoectively.
Usually at the NLO level ${\rm \overline{MS}}$-scheme is used, so we apply
$\Lambda = \Lambda_{\rm \overline{MS}}$ below.
%in the Eqs.~(\ref{an:NLO}) and (\ref{as:NLO}).

\section{Parton distributions and the structure function $F_2$
}

Here, for simplicity we consider only  the LO
%leading order (LO)
approximation\footnote{
The NLO results can be found in  \cite{Q2evo,HT}.}.
The structure function $F_2$  and PDFs $f_a$ $(a=q,g)$ have the form
\begin{eqnarray}
    F_2(x,Q^2) &=& e \, f_q(x,Q^2),~~
    f_a(x,Q^2)
%&=&
~=~ f_a^{+}(x,Q^2) + f_a^{-}(x,Q^2),
%~~(a=q,g)
\label{8a}
\end{eqnarray}
where
$e=(\sum_1^f e_i^2)/f$ is an average charge squared.

The small-$x$ asymptotic expressions for parton densities
%(PD)
$f^{\pm}_a$ look like
\begin{eqnarray}
    f^{+}_g(x,Q^2) &=& \biggl(A_g + \frac{4}{9} A_q \biggl)
        I_0(\sigma) \; e^{-\overline d_{+} s} + O(\rho),~~
    f^{+}_q(x,Q^2) ~=~
%&=&
\frac{f}{9} \frac{\rho I_1(\sigma)}{I_0(\sigma)} \, f^{+}_g(x,Q^2)
+ O(\rho),
%   \label{8.01} \\
\nonumber \\
    f^{-}_g(x,Q^2) &=& -\frac{4}{9} A_q e^{- d_{-} s} \, + \, O(x),~~
%   \label{8.00} \\
    f^{-}_q(x,Q^2)
%&=&
~=~ A_q e^{-d_{-}(1) s} \, + \, O(x),
    \label{8.02}
\end{eqnarray}
where $I_{\nu}$ ($\nu=0,1$)  are the modified Bessel
functions and $\sigma$ and $\rho$ can be found in
(\ref{intro:1}) when $p=0$. The coefficient $\hat{d}_+$ 
%was introduced in
(see eq. (\ref{intro:1a})) and
\begin{equation}
%\hat{d}_+ = - \frac{12}{\beta_0},~~~
\overline d_{+} = 1 + \frac{20f}{27\beta_0},~~~
d_{-} = \frac{16f}{27\beta_0}
\label{intro:1b}
\end{equation}
denote singular and regular parts of the anomalous dimensions
$d_{+}(n)$ and $d_{-}(n)$,
respectively, in the limit $n\to1$\footnote{
We denote the singular and regular parts of a given quantity $k(n)$ in the
limit $n\to1$ by $\hat k/(n-1)$ and $\overline k$, respectively.}.
Here $n$ is a variable in the Mellin space.

\begin{figure}[t]
\includegraphics[height=0.55\textheight,width=0.95\textwidth]{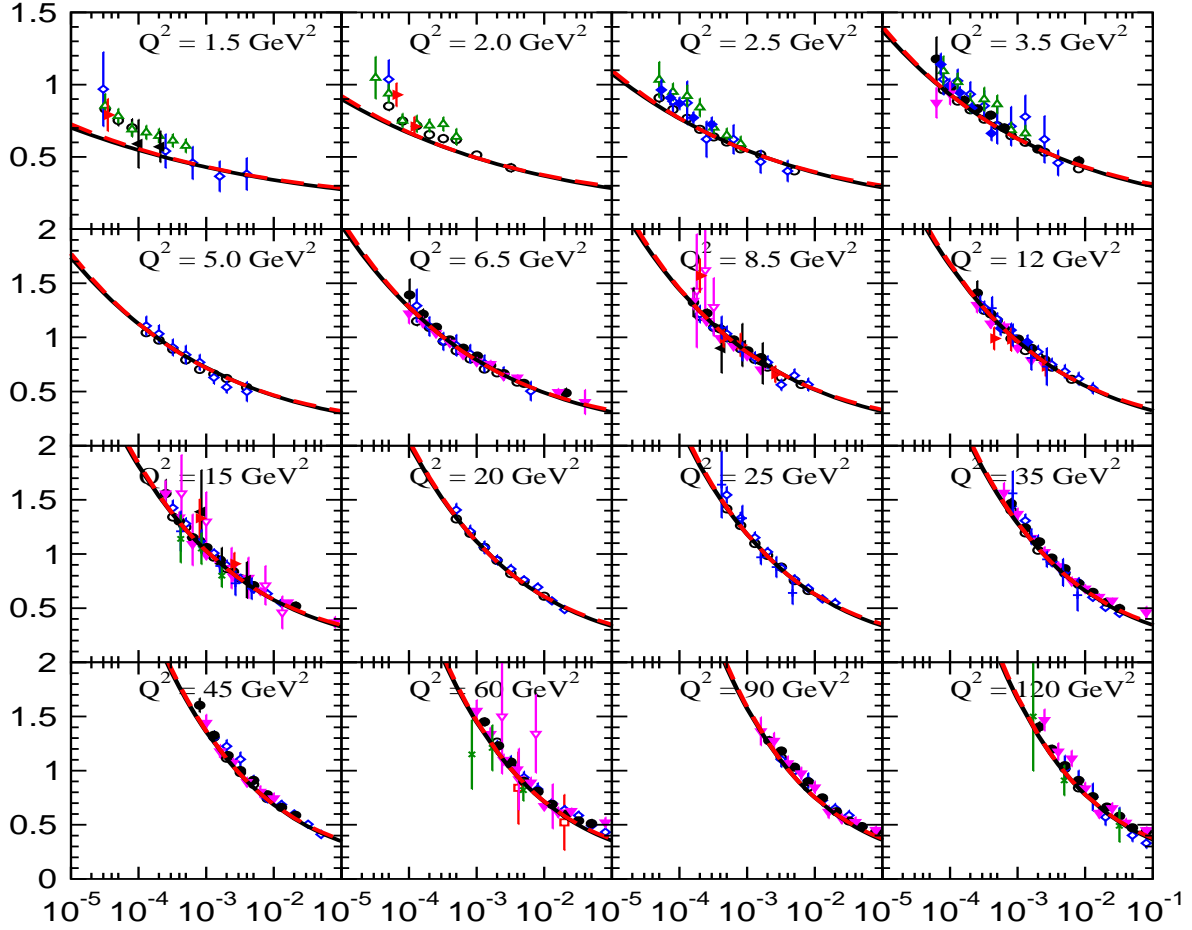}
%\vskip 1.3cm
\vspace{1.5cm}
\caption{$F_2(x,Q^2)$ as a function of $x$ for different $Q^2$ bins. 
The experimental points are from H1 \cite{H197} (open points) and ZEUS 
\cite{ZEUS01} (solid points) at 
$Q^2 \geq 1.5$ GeV$^2$.
The solid curve represents the NLO fit. The dashed curve (hardly 
distinguishable 
from the solid one) represents the LO fit.}
\label{fig1}
\end{figure}

\begin{figure}[t]
\includegraphics[height=0.55\textheight,width=0.95\textwidth]{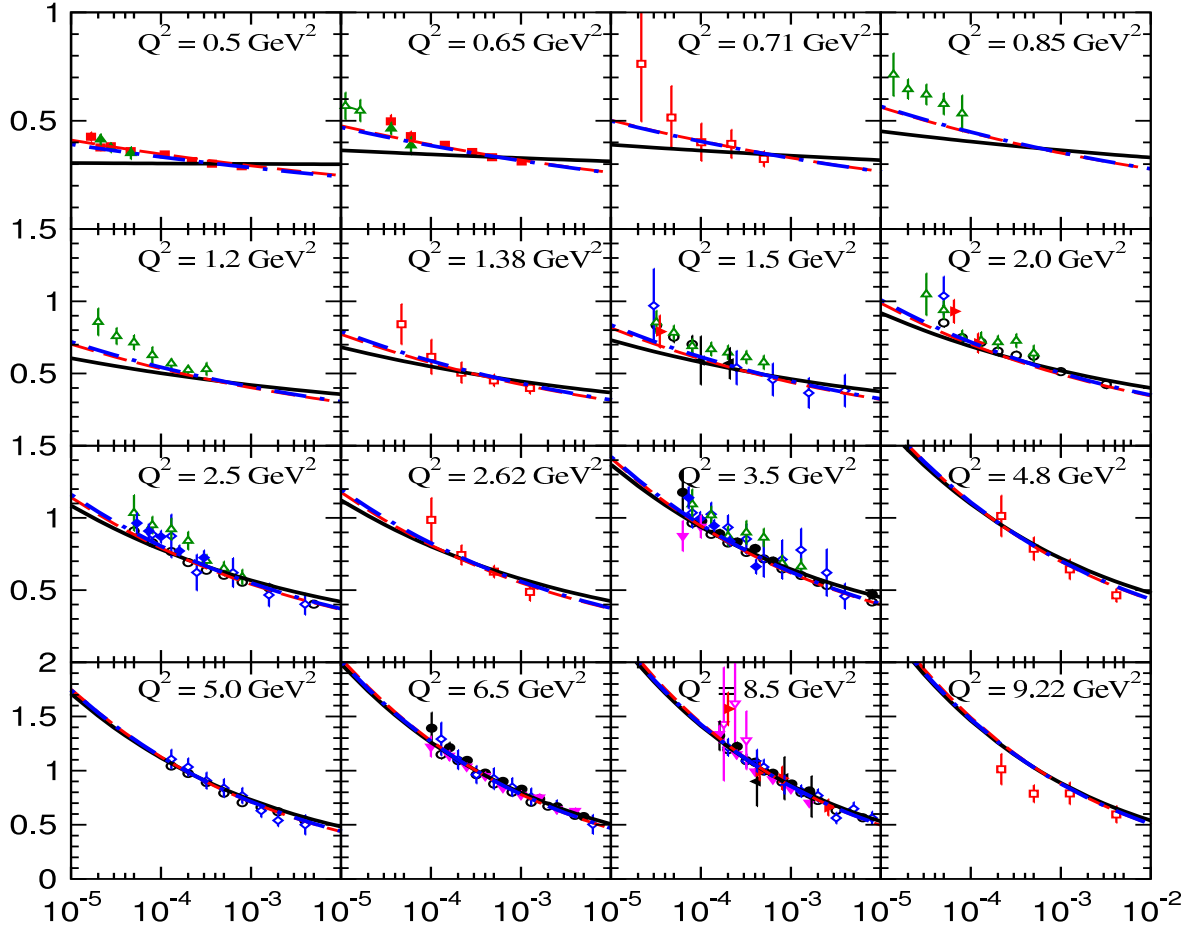}
%\vskip 1.5cm
\caption{ $x$ dependence of $F_2(x, Q^2)$ in bins of $Q^2$. The experimental data from H1 (open points) 
and ZEUS (solid points) are compared with the NLO fits for $Q^2\geq 0.5$ GeV$^2$
implemented with the canonical (solid lines), frozen (dot-dashed lines), and analytic (dashed lines) 
versions of the strong-coupling constant. For comparison, also the results
obtained in Ref. \cite{HT} through a fit based on the renormalon model of higher-twist terms are shown 
(dotted lines).}
\label{fig3}
\end{figure}

%\begin{\Large}
\begin{table}
\caption{
%\label{Tab:H1+ZEUS:96/97}\sffamily
The result of the LO and NLO fits to H1 
%(1996/97) \protect\cite{Adloff:1999}
and ZEUS 
%(1996/97) \protect\cite{Chekanov:2001} 
data  for different low
$Q^2$ cuts.  In the fits $f$ is fixed to 4 flavors.
}
\centering
\footnotesize
%\small
%\large
\vspace{0.3cm}
%\begin{ruledtabular}
\begin{tabular}{|l||c|c|c||r|} \hline \hline
& $A_g$ & $A_q$ & $Q_0^2~[{\rm GeV}^2]$ &
 $\chi^2 / n.o.p.$~ \\
\hline\hline
~$Q^2 \geq 1.5 {\rm GeV}^2 $  &&&& \\
 LO & 0.784$\pm$.016 & 0.801$\pm$.019 & 0.304$\pm$.003 & 754/609 \\
 LO$\&$an. & 0.932$\pm$.017 & 0.707$\pm$.020 & 0.339$\pm$.003 & 632/609  \\
  LO$\&$fr. & 1.022$\pm$.018 & 0.650$\pm$.020 & 0.356$\pm$.003 & 547/609   \\
\hline
 NLO & -0.200$\pm$.011 & 0.903$\pm$.021 & 0.495$\pm$.006 & 798/609 \\
 NLO$\&$an. & 0.310$\pm$.013 & 0.640$\pm$.022 & 0.702$\pm$.008 & 655/609  \\
  NLO$\&$fr. & 0.180$\pm$.012 & 0.780$\pm$.022 & 0.661$\pm$.007 & 669/609   \\
\hline\hline
~$Q^2 \geq 0.5 {\rm GeV}^2 $  &&&& \\
 LO & 0.641$\pm$.010 & 0.937$\pm$.012 & 0.295$\pm$.003 & 1090/662 \\
 LO$\&$an. & 0.846$\pm$.010 & 0.771$\pm$.013 & 0.328$\pm$.003 & 803/662  \\
  LO$\&$fr. & 1.127$\pm$.011 & 0.534$\pm$.015 & 0.358$\pm$.003 & 679/662   \\
\hline
 NLO & -0.192$\pm$.006 & 1.087$\pm$.012 & 0.478$\pm$.006 & 
%{\color{red} 
1229/662
%} 
\\
 NLO$\&$an. & 0.281$\pm$.008 & 0.634$\pm$.016 & 0.680$\pm$.007 & 
%{\color{red} 
633/662
%}  
\\
  NLO$\&$fr. & 0.205$\pm$.007 & 0.650$\pm$.016 & 0.589$\pm$.006 & 
%{\color{red} 
670/662
%}   
\\
\hline \hline
%\normale
\end{tabular}
%\end{ruledtabular}
\end{table}

\begin{figure}[htb]
%\begin{center}
%\includegraphics[height=4.9in,width=6.5in]{Fig5.ps}
\includegraphics[height=4.5in,width=5.8in]{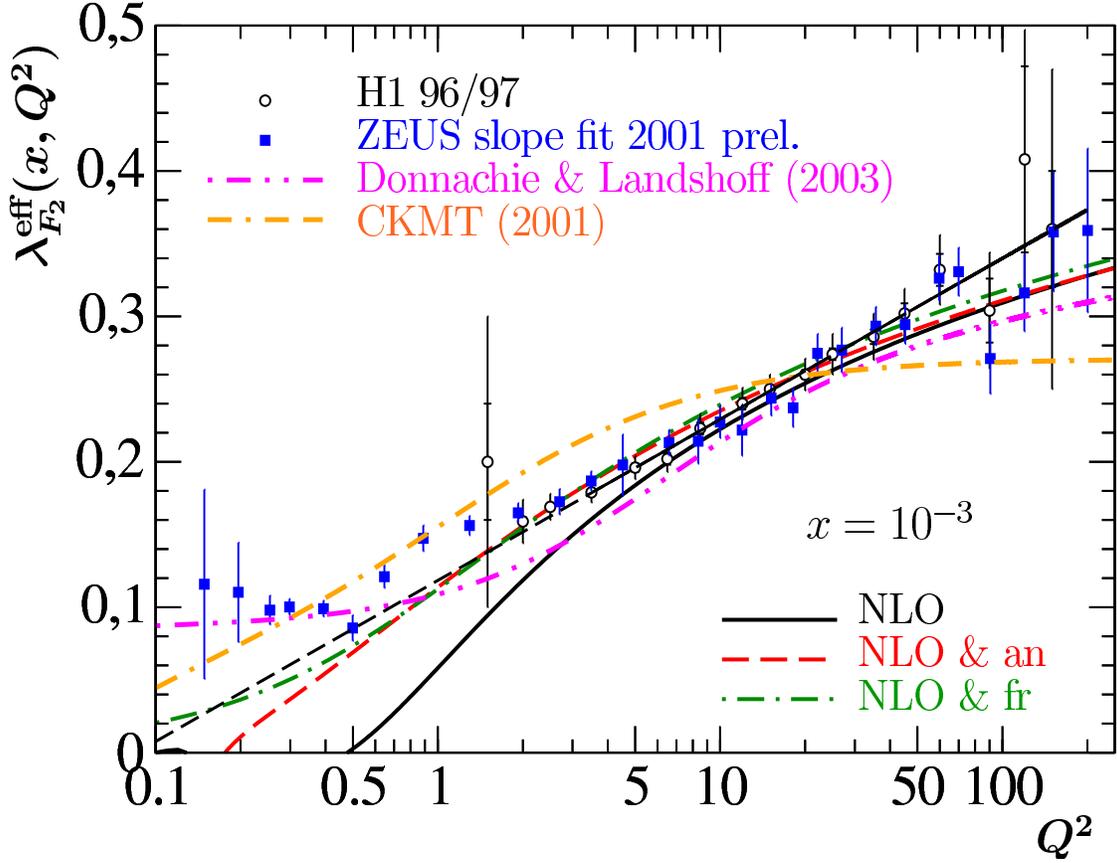}
%\includegraphics[width=0.9\textwidth]{Fig5.ps}
%\includegraphics[height=0.9\textheight,width=0.9\textwidth]{Fig5.ps}
%\end{center}
%\vskip 1cm
\caption{
$Q^2$ dependence of $\lambda^{\rm eff}_{\rm F_2}(x,Q^2)$
for an average small-$x$ value of $x = 10^{-3}$. The
experimental data from H1 (open points) and ZEUS (solid points) are compared
with the NLO fits for $Q^2 \geq 0.5$ GeV$^2$ implemented with the canonical (solid line),
frozen (dot-dashed line), and analytic (dashed line) versions of the strong-coupling
constant. The linear rise of $\lambda^{\rm eff}_{\rm F_2}(x,Q^2)$
 with $\ln Q^2$ 
%as described by Eq. (2) 
is indicated
by the straight dashed line. For comparison, also the results obtained in the
phenomenological models by Capella et al.
%Kaidalov et al. 
%[47] 
\cite{CaKaMeTTV} (dash-dash-dotted line) and by
Donnachie and Landshoff \cite{Donnachie:2003cs} 
(dot-dot-dashed line) are shown.}
\label{fig5}
\end{figure}

\begin{figure}[t]
\centering
%\vskip 0.5cm
%\includegraphics[width=.75\hsize]{F2sm-t2A.ps}
%\includegraphics[width=.75\hsize]{smallx.eps}
%\includegraphics[width=.95\hsize]{smallx.eps}
\includegraphics[height=0.75\textheight,width=1.05\hsize]{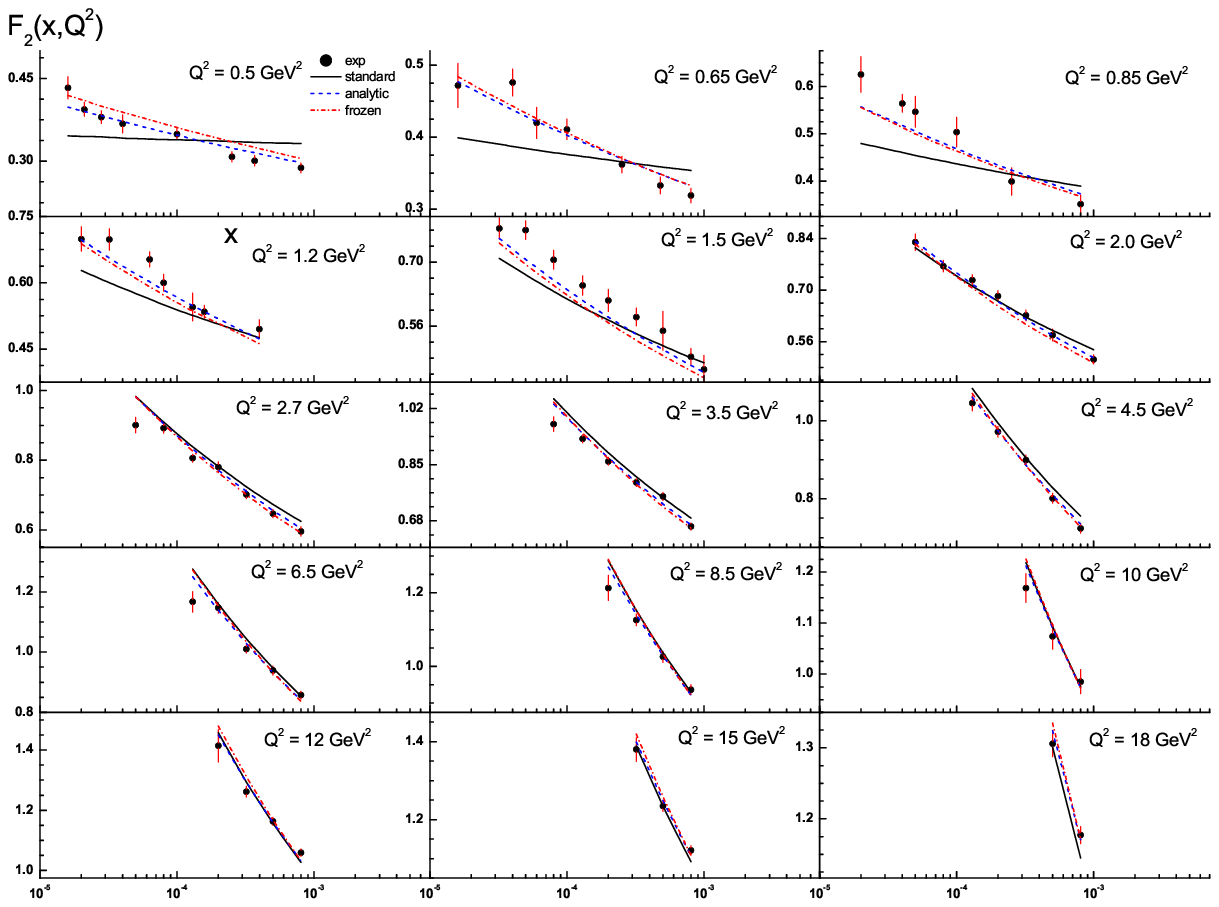}
%\includegraphics[height=0.75\textheight]{smallx.eps}
%\vskip -0.3cm
\caption{$x$ dependence of $F_2(x,Q^2)$ in bins of $Q^2$.
The combined experimental data from H1 and ZEUS Collaborations
\cite{Aaron:2009aa} are
compared with the NLO fits for $Q^2\geq0.5$~GeV$^2$ implemented with the
standard (solid lines), frozen (dot-dashed lines), and analytic (dashed lines)
versions of the strong coupling constant.}
\label{fig:F1}
\end{figure}

\section{Effective slopes}

Contrary to the approach in  \cite{BF1}-\cite{HT}
%\cite{BF1,Munich,Q2evo,HT}
%\cite{BF1}-\cite{HT},
%On the other hand, 
various groups have been able to fit
the available data 
%({\it separately at low and high $Q^2$ values})
using a hard input at small $x$: 
$x^{-\lambda},~\lambda >0$ with different $\lambda$ values at low and high 
$Q^2$ (see \cite{LoYn}-\cite{DeJePa}).
%\cite{LoYn,DoLa,Abramo,YF93,FKR,CaKaMeTTV}).
Such results are well-known at low $Q^2$ values \cite{DoLa}. 
At large $Q^2$ values, for 
%In some sense, it is not very surprising, because 
the modern HERA data 
it is also not very surprising, because improssible to
%they cannot 
distinguish between the behavior
based on a steep input parton parameterization,
at quite large $Q^2$, and the
steep form acquired after the dynamical evolution from a flat initial
condition at quite low $Q^2$ values.

As it has been mentioned above and shown in \cite{Q2evo,HT,KoPa02},
the behavior of parton densities and $F_2$ given in the Bessel-like form 
by generalized DAS approach
%Eqs. (\ref{9.10})-(\ref{9}) 
can mimic a power law shape
over a limited region of $x$ and $Q^2$
 \begin{eqnarray}
f_a(x,Q^2) \sim x^{-\lambda^{\rm eff}_a(x,Q^2)}
 ~\mbox{ and }~
F_2(x,Q^2) \sim x^{-\lambda^{\rm eff}_{\rm F_2}(x,Q^2)}.
\nonumber    \end{eqnarray}

The effective slopes $\lambda^{\rm eff}_a(x,Q^2)$ and $\lambda^{\rm eff}_{\rm F_2}(x,Q^2)$
have the form:
 \begin{eqnarray}
\lambda^{\rm eff}_g(x,Q^2) ~=~
%&=& 
\frac{f^+_g(x,Q^2)}{f_g(x,Q^2)} \,
\rho \, \frac{I_1(\sigma)}{I_0(\sigma)},~~~~
%\nonumber\\
\lambda^{\rm eff}_{\rm F_2}(x,Q^2) ~=~
\lambda^{\rm eff}_q(x,Q^2) ~=~ \frac{f^+_q(x,Q^2)}{f_q(x,Q^2)} \,
\rho \, \frac{ I_2(\sigma) 
%(1- 20 a_s(Q^2)) + 20 a_s(Q^2) \tilde I_1(\sigma)/\rho
}{I_1(\sigma) 
%(1- 20 a_s(Q^2)) + 20 a_s(Q^2) \tilde I_0(\sigma)/\rho
}.
%\nonumber
\label{10.1}
\end{eqnarray}
%where the exact form of parton densities can be found in \cite{Q2evo,HT}.

The effective slopes $\lambda^{\rm eff}_a $ and 
$\lambda^{\rm eff}_{\rm F_2}$ depend on the magnitudes $A_a$ of the initial PDFs
and also on the chosen input values of $Q^2_0$ and $\Lambda $.
To compare with the experimental data it is necessary the exact expressions
(\ref{10.1}), but for qualitative analysis it is better to use an 
approximation.
%\subsection{Asymptotic form of the effective slopes}
%
%
At quite 
large values of $Q^2$, 
where the ``$-$'' component is negligible,
the dependence on the initial PDFs disappears, having
in this case for the asymptotic behavior the following 
expressions:
 \begin{eqnarray}
\lambda^{\rm eff,as}_g(x,Q^2) ~=~
%&=& 
\rho\, \frac{I_1(\sigma)}{I_0(\sigma)} \approx \rho - 
\frac{1}{4\ln{(1/x)}}, ~~~
%\nonumber \\
%\label{11.1a} \\
\lambda^{\rm eff,as}_{\rm F_2}(x,Q^2) ~=~
%&=&
\lambda^{\rm eff,as}_q(x,Q^2) ~=~ 
\rho \frac{I_2(\sigma) 
%(1- 20 a_s(Q^2)) + 20 a_s(Q^2)\tilde  I_1(\sigma)/\rho
}{I_1(\sigma) 
%(1- 20 a_s(Q^2)) + 20 a_s(Q^2)\tilde  I_0(\sigma)/\rho
}
% \nonumber \\
%&=& \rho\, \frac{\tilde I_2(\sigma)}{\tilde I_1(\sigma)} 
% + 20 a_s(Q^2)\left( 1- \frac{\tilde  I_0(\sigma) 
%\tilde  I_2(\sigma)}{\tilde  I_1^2(\sigma)} \right)
\approx 
 \rho - \frac{3}{4\ln{(1/x)}} 
%+  \frac{10a_s(Q^2)}{ \rho \ln{(1/x)}}, 
%
% \nonumber \\
\label{11.1b}
\end{eqnarray}
where the symbol $\approx $ marks the approximation obtained in the  expansion
of the 
%usual and 
modified Bessel functions.
% in (\ref{8.01}).

\section{``Frozen'' and analytic  coupling constants}

In order to improve an agreement at low $Q^2$ values,
the QCD coupland
%coupling constant 
is modified in the infrared region.
We considered  \cite{Cvetic:2009kw}
two modifications that effectively increase the argument of 
the coupling constant at low $Q^2$ values (see \cite{DoShi}).

In the first case, which is more phenomenological, we introduce freezing
of the coupling constant by changing its argument $Q^2 \to Q^2 + M^2_{\rho}$,
where $M_{\rho}$ is the $\rho $-meson mass (see \cite{Greco}). 
Thus, in the above
formulae 
%of Sec. 2 
we have to carry out the following replacement:
\begin{equation}
 a_s(Q^2) \to a_{\rm fr}(Q^2) \equiv a_s(Q^2 + M^2_{\rho})
\label{Intro:2}
\end{equation}

The second possibility follows the Shirkov--Solovtsov idea
\cite{ShiSo}
concerning the analyticity of the coupling constant that leads
to additional power dependence of the latter.
Then, in the above formulae 
%of the previous section
%and \ref{Sec:3}
the coupling constant $a_s(Q^2)$ should be replaced as follows:
\begin{eqnarray}
 a^{\rm LO}_{\rm an}(Q^2) \, = \, a^{\rm LO}_s(Q^2) - \frac{1}{\beta_0}
 \frac{\Lambda^2_{\rm LO}}{Q^2 - \Lambda^2_{\rm LO}},~~~~~~
a_{\rm an}(Q^2) \, = \, a_s(Q^2) - \frac{1}{2\beta_0}
\frac{\Lambda^2}{Q^2 - \Lambda^2}
+ \ldots \, ,
\label{an:LO}
\end{eqnarray}
in the LO and NLO approximations, respectively. 
Here the the symbol $\ldots$ stands
for the terms that provide negligible contributions when $Q^2 \geq 1$ GeV \cite{ShiSo}.
Note 
%here 
that the perturbative coupling constant $a_s(Q^2)$
is different in the LO
%leading-order (LO)
and NLO approximations (see eq. (\ref{as:LO}) 
%and (\ref{as:NLO}) 
above).

\section{Comparison with experimental data
%Results of the fits
} \indent

Using the results of previous section we have
analyzed  \cite{Q2evo,HT,Cvetic:2009kw}
HERA data for $F_2$ and the slope $\partial \ln F_2/\partial \ln (1/x)$
at small $x$ from the H1 and ZEUS Collaborations \cite{H197}-\cite{DIS02}.
%\cite{H197,ZEUS01,Surrow,H1slo,DIS02}.
%
In order to keep the analysis as simple as possible,
we fix $f=4$ and $\alpha_s(M^2_Z)=0.1166 $ (i.e., $\Lambda^{(4)} = 284$ MeV) in agreement
with the more recent ZEUS results \cite{ZEUS01}.

%\subsection{H1 and ZEUS data for $F_2$}

As it is possible to see in Fig. 1 (see also \cite{Q2evo,HT}), the twist-two
approximation is reasonable at $Q^2 \geq 2$ GeV$^2$. 
Moreover, 
 the results of fits in \cite{HT} have an important property: they are
very similar in LO and NLO approximations of perturbation theory.
The similarity is related to the fact that the small-$x$ asymptotics of 
the NLO corrections
are usually large and negative (see, for example, $\alpha_s$-corrections 
\cite{FaLi,KoLi} to Balitsky--Fadin--Kuraev--Lipatov (BFKL)
%BFKL 
kernel \cite{BFKL}\footnote{It seems that it is a property of 
any processes in which gluons,
but not quarks play a basic role.}).
% and 
Then, the LO form $\sim \alpha_s(Q^2)$ for
some observable and the NLO one 
$\sim \alpha_s(Q^2) (1-K\alpha_s(Q^2)) $
with a large value of $K$ are similar, because 
%usually 
$\Lambda \gg
\Lambda_{\rm LO}$\footnote{The equality of
%similarity between 
$\alpha_s(M_Z^2)$ at LO and NLO approximations,
%and $\alpha^{\rm LO}_s(M_Z^2)$,
where $M_Z$ is the $Z$-boson mass, relates $\Lambda$ and $\Lambda_{\rm LO}$:
$\Lambda^{(4)} = 284$ MeV (as in \cite{ZEUS01}) corresponds to 
$\Lambda_{\rm LO} = 112$ MeV (see \cite{HT}).}
and, thus, $\alpha_s(Q^2)$ at LO is considerably smaller  then 
$\alpha_s(Q^2)$ at NLO  for HERA $Q^2$ values.

In other words, performing some resummation procedure (such as Grunberg's 
effective-charge method \cite{Grunberg}), one can see that the
%the NLO form 
results up to NLO approximation may
%can 
be represented as $\sim \alpha_s(Q^2_{\rm eff})$,
where $Q^2_{\rm eff} \gg Q^2$. 
Indeed, from 
different studies
\cite{DoShi,bfklp,Andersson},
it is well known that at small-$x$ values the effective
argument of the coupling constant is higher then $Q^2$.

At smaller $Q^2$, some
modification of the twist-two approximation should be considered. In Ref.
%the recent article
\cite{HT} we have added the higher twist corrections.
For renormalon model of higher twists, we
have found a good
agreement with experimental data at essentially lower $Q^2$ values:
$Q^2 \geq 0.5$ GeV$^2$ (see Figs. 2 and 3 in \cite{HT}), but we have added 
4 additional parameters:
amplitudes of twist-4 and twist-6 corrections to quark and gluon densities.

%Here, to 
To improve the agreement at small $Q^2$ values without additional parameters,
we modified \cite{Cvetic:2009kw}
the QCD coupling constant.
% (see \cite{Cvetic:2009kw}).
We considered two modifications: analytic and frozen coupling constants, 
which effectively increase the argument of the coupling constant 
at small $Q^2$ values (in agreement with \cite{DoShi,bfklp,Andersson}).

%\subsection{H1 and ZEUS data for the slope $\lambda^{\rm eff}_{\rm F_2}$}

Figure~2 and Table 1 show  a strong improvement of the agreement with experimental data
for $F_2$ (almost 2 times!). Similar results can be seen also in 
Fig. 3 for
%shows 
the experimental data for $\lambda_{F_2}^{\rm eff}(x,Q^2)$
at $x\sim 10^{-3}$, which represents an average of the $x$-values of HERA experimental 
data. 
Note that
%timplementing 
the ``frozen'' and analytic coupling constants
$\alpha_{\rm fr}(Q^2)$ and $\alpha_{\rm an}(Q^2)$, 
%respectively, which in turn 
lead to
very close results (see also \cite{KoLiZo,Kotikov:2010bm}).

Indeed, the fits for $F_2(x,Q^2)$ in \cite{HT}
yielded
$Q^2_0 \approx 0.5$--$0.8$~GeV$^2$.
So, initially we had $\lambda^{\rm eff}_{F_2}(x,Q^2_0)=0$,
as suggested by Eq.~(\ref{1}). The replacements of Eqs.~(\ref{Intro:2}) and
 (\ref{an:LO}) 
%and (\ref{an:NLO}) 
modify the value of $\lambda^{\rm eff}_{F_2}(x,Q^2_0)$. 
%So, for 
For the  
``frozen'' and analytic coupling constants 
$\alpha_{\rm fr}(Q^2)$ and $\alpha_{\rm an}(Q^2)$,
%coupling constant $a_{fr}(Q^2)$ 
the value of
$\lambda^{\rm eff}_{F_2}(x,Q^2_0)$ is nonzero 
%now 
and the slopes are
%is 
quite close to the experimental data at $Q^2 \approx 0.5$~GeV$^2$.
Nevertheless, for $Q^2 \leq 0.5$~GeV$^2$, there is still some disagreement with
the data for the slope $\lambda^{\rm eff}_{F_2}(x,Q^2)$, which needs 
additional investigation.
Note that at $Q^2 \geq 0.5$ GeV$^2$ our results for 
$\lambda^{\rm eff}_{F_2}(x,Q^2)$
are even better the results of 
phenomenological models \cite{CaKaMeTTV,Donnachie:2003cs}.

%\subsection{Combined H1$\&$ZEUS data for $F_2$}

%\begin{\Large}
\begin{table}
\caption{
%\label{Tab:H1+ZEUS:96/97}\sffamily
The results of LO and NLO fits to  H1 $\&$ ZEUS data
%for $F_2$
\cite{Aaron:2009aa},
%H1
%%(1996/97) \protect\cite{Adloff:1999}
%and ZEUS
%%(1996/97) \protect\cite{Chekanov:2001}
with various lower cuts on $Q^2$; in the fits
the number of flavors $f$ is fixed to 4.
}
\centering
\footnotesize
%\small
%\large
\vspace{0.3cm}
%\begin{ruledtabular}
\begin{tabular}{|l||c|c|c||r|} \hline \hline
& $A_g$ & $A_q$ & $Q_0^2~[{\rm GeV}^2]$ &
 $\chi^2 / n.d.f.$~ \\
\hline\hline
~$Q^2 \geq 5 {\rm GeV}^2 $  &&&& \\
 LO & 0.623$\pm$0.055 & 1.204$\pm$0.093 & 0.437$\pm$0.022 & 1.00 \\
 LO$\&$an. & 0.796$\pm$0.059 & 1.103$\pm$0.095 & 0.494$\pm$0.024 & 0.85  \\
  LO$\&$fr. & 0.782$\pm$0.058 & 1.110$\pm$0.094 & 0.485$\pm$0.024 & 0.82   \\
\hline
 NLO & -0.252$\pm$0.041 & 1.335$\pm$0.100 & 0.700$\pm$0.044 & 1.05 \\
 NLO$\&$an. & 0.102$\pm$0.046 & 1.029$\pm$0.106 & 1.017$\pm$0.060 & 0.74  \\
  NLO$\&$fr. & -0.132$\pm$0.043 & 1.219$\pm$0.102 & 0.793$\pm$0.049 & 0.86   \\
\hline\hline
~$Q^2 \geq 3.5 {\rm GeV}^2 $  &&&& \\
 LO & 0.542$\pm$0.028 & 1.089$\pm$0.055 & 0.369$\pm$0.011 & 1.73 \\
 LO$\&$an. & 0.758$\pm$0.031 & 0.962$\pm$0.056 & 0.433$\pm$0.013 & 1.32  \\
  LO$\&$fr. & 0.775$\pm$0.031 & 0.950$\pm$0.056 & 0.432$\pm$0.013 & 1.23   \\
\hline
 NLO & -0.310$\pm$0.021 & 1.246$\pm$0.058 & 0.556$\pm$0.023 & 1.82 \\
 NLO$\&$an. & 0.116$\pm$0.024 & 0.867$\pm$0.064 & 0.909$\pm$0.330 & 1.04  \\
  NLO$\&$fr. & -0.135$\pm$0.022 & 1.067$\pm$0.061 & 0.678$\pm$0.026 & 1.27 \\
\hline \hline
~$Q^2 \geq 2.5 {\rm GeV}^2 $  &&&& \\
 LO & 0.526$\pm$0.023 & 1.049$\pm$0.045 & 0.352$\pm$0.009 & 1.87 \\
 LO$\&$an. & 0.761$\pm$0.025 & 0.919$\pm$0.046 & 0.422$\pm$0.010 & 1.38  \\
  LO$\&$fr. & 0.794$\pm$0.025 & 0.900$\pm$0.047 & 0.425$\pm$0.010 & 1.30   \\
\hline
 NLO & -0.322$\pm$0.017 & 1.212$\pm$0.048 & 0.517$\pm$0.018 & 2.00 \\
 NLO$\&$an. & 0.132$\pm$0.020 & 0.825$\pm$0.053 & 0.898$\pm$0.026 & 1.09  \\
  NLO$\&$fr. & -0.123$\pm$0.018 & 1.016$\pm$0.051 & 0.658$\pm$0.021 & 1.31   \\
\hline\hline
~$Q^2 \geq 0.5 {\rm GeV}^2 $  &&&& \\
 LO & 0.366$\pm$0.011 & 1.052$\pm$0.016 & 0.295$\pm$0.005 & 5.74 \\
 LO$\&$an. & 0.665$\pm$0.012 & 0.804$\pm$0.019 & 0.356$\pm$0.006 & 3.13  \\
  LO$\&$fr. & 0.874$\pm$0.012 & 0.575$\pm$0.021 & 0.368$\pm$0.006 & 2.96   \\
\hline
 NLO & -0.443$\pm$0.008 & 1.260$\pm$0.012 & 0.387$\pm$0.010 & 6.62 \\
 NLO$\&$an. & 0.121$\pm$0.008 & 0.656$\pm$0.024 & 0.764$\pm$0.015 & 1.84  \\
  NLO$\&$fr. & -0.071$\pm$0.007 & 0.712$\pm$0.023 & 0.529$\pm$0.011 & 2.79 \\
\hline \hline
%\hline
%\normale
\end{tabular}
%\end{ruledtabular}
\end{table}

At the next step we considered \cite{Kotikov:2012sm} the
%By using the results of the previous section we have analyzed 
combined H1$\&$ZEUS data for $F_2$ \cite{Aaron:2009aa}.
As can be seen from Fig.~4 and Table~2,
%(see also \cite{Q2evo,HT}),
the twist-two approximation is reasonable for $Q^2 \geq 2$ GeV$^2$.
At lower $Q^2$ we observe that the fits in the cases with ``frozen'' and
analytic strong coupling constants are very similar
(see also \cite{KoLiZo,Cvetic:2009kw,Kotikov:2010bm}) and describe the data in
the low $Q^2$ region significantly better than the standard fit.
Nevertheless, for $Q^2 \leq 1.5$~GeV$^2$
%in the case of $\lambda^{\rm eff}_{F_2}(x,Q^2_0)$,
there is still some disagreement with
the data, which needs to be additionally studied.
In particular,  the BFKL
%Balitsky--Fadin--Kuraev--Lipatov (BFKL)
resummation \cite{BFKL} may be important here \cite{Kowalski:2012ur}.
It can be added in the generalized DAS approach according to the discussion
in Ref. \cite{KoBaldin}.

%\vspace{-0.3cm}
\section{Conclusions} \indent

We have shown
%studied 
the $Q^2$-dependence of the structure functions $F_2$ and 
%$F_2^{cc}$ and of 
the slope 
$\lambda^{\rm eff}_{F_2}=\partial \ln F_2/\partial \ln (1/x)$ at 
small-$x$ values in the 
framework of perturbative QCD. Our twist-two 
results are in a very good agreement with 
%new 
precise HERA data for $Q^2 \geq 2
%\div 4
$~GeV$^2$,
where perturbative theory 
%can be 
is applicable.
%The application of 
Using the ``frozen'' and analytic coupling constants 
$\alpha_{\rm fr}(Q^2)$
and $\alpha_{\rm an}(Q^2)$ improves
an agreement with the recent HERA data \cite{Surrow,H1slo,DIS02}
for the slope $\lambda^{\rm eff}_{F_2}(x,Q^2)$ for small $Q^2$ values,
$Q^2 \geq 0.5$~GeV$^2$.

As the next spep, we are going to adopt
%plan to apply 
the Grunberg approach \cite{Grunberg}
together with
the ``frozen'' and analytic modifications of the
strong coupling constant for analyse of the combined H1$\&$ZEUS data 
for $F_2$ \cite{Aaron:2009aa}. The similar study has been done recently 
\cite{Kotikov:2012eq} for 
%analising of 
experimental data of the Bjorken 
sum rule.

%%%%%%%%%%%%%%%%%%%%%%%%%%%%%%%%%%%%%%%%%%%%%%%%
%% BACKMATTER
%%%%%%%%%%%%%%%%%%%%%%%%%%%%%%%%%%%%%%%%%%%%%%%%

\begin{theacknowledgments}
This work was supported by 
RFBR grant 13-02-01060-a.
%Author 
A.V.K. 
thanks the Organizing Committee of II Russian-Spanish Congress
for invitation and support.

\end{theacknowledgments}



\bibliographystyle{aipproc}   % if natbib is available

\bibliography{sample}

%

%%%%%%%%%%%%%%%%%%%%%%%%%%%%%%%%%%%%%%%%%%%
%% The following lines show an example how to produce a bibliography
%% without the help of the BibTeX program. This could be used instead
%% of the above.
%%%%%%%%%%%%%%%%%%%%%%%%%%%%%%%%%%%%%%%%%%%

\end{document}